\documentstyle[prl,aps,epsf,floats,twocolumn]{revtex}
\begin{document}
\twocolumn[\hsize\textwidth\columnwidth\hsize\csname
@twocolumnfalse\endcsname
\title{Supercluster Magnetic Fields and Anisotropy of Cosmic Rays
above $10^{19}\,$eV}
\author{Martin Lemoine}
\address{DARC, UMR--8629, CNRS,
Observatoire de Paris-Meudon, F-92195 Meudon C\'edex, France}
\author{G\"unter Sigl}
\address{DARC, UMR--8629, CNRS, Observatoire de Paris-Meudon, F-92195
Meudon C\'{e}dex, France\\
and\\
Department of Astronomy \& Astrophysics,
Enrico Fermi Institute, The University of Chicago, Chicago, 
IL~~60637-1433}
\author{Peter Biermann}
\address{Max-Planck Institute for Radioastronomy,
D-53010 Bonn, Germany}

\date{\today}
\maketitle

\begin{abstract} We predict energy spectra and angular distributions of
nucleons above $10^{19}\,$eV that originate from sources distributed in
the Local Supercluster, which is also supposed to contain a large scale
magnetic field of strength $\sim0.05-0.5\,\mu$G. We show that this
model can explain all present-day features of ultra-high energy cosmic
rays, at least for field strengths close to $0.5\,\mu$G. The
large-scale anisotropy and the clustering predicted by this scenario
will allow strong discrimination against other models with next
generation experiments.

PACS numbers: 98.70.Sa, 98.62.En

\end{abstract}
\vskip2.2pc]


Ryu \& Biermann~\cite{RKB98} have recently argued that the
observational upper limit on the strength $B_{\rm rms}$ of an
extra-galactic magnetic field (EGMF), obtained from Faraday rotation
observations of distant sources~\cite{K94,V97}, reads:  $B_{\rm
rms}\lesssim 1\,\mu$G, for fields contained inside the cosmological
large-scale structure, such as the Local Supercluster. Such a strong
magnetic field would have profound consequences on the propagation of
charged ultra-high energy cosmic rays (UHECRs) with energy
$E\gtrsim10\,$EeV
($1\,$EeV$\equiv10^{18}\,$eV)~\cite{WW79,SLB99,BO99,MT98}. In
particular, for $B_{\rm rms}\sim0.1\,\mu$G, charged UHECRs with
energies up to $\sim100\,$EeV would diffuse, while UHECRs of higher
energies would propagate in nearly straight lines. This would allow to
reproduce very nicely the observed energy spectrum of UHECRs for a
single injection spectrum $\propto E^{-2.4}$~\cite{SLB99,BO99}.
Moreover, the associated angular deflection might explain why no
astrophysical counterpart within $\simeq50\,$Mpc could be associated to
the highest energy events. The existence of a magnetic field of
strength $\sim0.1\mu\,$G could thus reconciliate models in which the
UHECRs are protons accelerated in conventional astrophysical
sources~\cite{NMA95}, with present-day observations.

The most recent results of the AGASA experiment~\cite{AGASA} provide
tight constraints on these ideas and more generally on any scenario of
UHECR origin. In particular, this experiment now reports the detection
of 7 events above $100\,$EeV, scattered across half the sky, with no
obvious association with the Supergalactic plane. A naive one-source
model, as considered in Refs.~\cite{SLB99,BO99}, which predicts an
angular size of the image $\sim15^\circ$ for $E\gtrsim100\,$EeV, is
thus excluded. However, whereas AGASA found no significant large scale
anisotropy, the data indicate significant small scale clustering. In
this letter we demonstrate that both the observed spectrum and the
angular distribution can be explained by a diffuse distribution of
sources with a density proportional to the matter density in the Local
Supercluster, provided this structure is permeated by magnetic fields
of strength $B_{\rm rms}\sim0.5\,\mu$G, with power concentrated on
$\sim\,$Mpc scales.  In this scenario, the large scale isotropy
observed by AGASA is explained by diffusion, whereas the small scale
clustering is due to magnetic focusing in the magnetic field
structure.  We discuss future observational tests of this scenario
against other models, especially with regard to the strong increase in
UHECR statistics anticipated from the Pierre Auger
Observatory~\cite{C92}.

{\it Energy spectrum and angular images.} During their propagation,
UHECR nucleons lose energy by pion production (for $E\gtrsim50\,$EeV),
and pair production (protons only), on the cosmic microwave
background. Charged UHECRs also acquire stochastic deflection, and
hence time delays with respect to straight line propagation, in random
magnetic fields. Detailed predictions for the energy spectrum and
angular distribution of UHECRs propagating in a magnetic field can only
be made through numerical Monte-Carlo
simulations~\cite{SLB99,SL98,LSO97}, in order to take into account
stochastic energy losses, stochastic deflection, and effects of an
anisotropic geometry. In order to correctly reproduce the small scale
angular distribution of events in a given realization of the magnetic
field, one needs to accomodate a very small solid angle, that
represents the detector as seen from the source, with a reasonable
consumption of CPU time. Each of the simulations presented below
typically requires one to several weeks of CPU time on DEC~ALPHA~500
computers. Details on our method can be found in
Refs.~\cite{SLB99,SL98,LSO97}.

We assume that the sources of UHECRs are distributed according to the
matter density in the Local Supercluster, following a pancake profile
with scale height of $5\,$Mpc and scale length of $20\,$Mpc; the
observer is located $20\,$Mpc away from the center of the density
profile (the Virgo cluster), and within $2\,$Mpc from the middle
plane.  Furthermore, we assume that no sources are present within 2 Mpc
from the observer.  This represents a reasonable modeling of our
location in the Local Supercluster and of its shape.  We also assume
that the Local Supercluster is permeated by a random magnetic field of
strength $0.05-0.5\,\mu$G, whose power spectrum follows a Kolmogorov
law $\langle B(k)^2\rangle\propto k^{n_B}$, with $n_B=-11/3$. The
largest eddy, defined as the scale over which the phase of the magnetic
field changes by $2\pi$, is $10\,$Mpc, and the corresponding smallest
eddy is $1\,$Mpc. This latter value is limited by resolution, but it
does not influence our results, as long as the power is concentrated on
the largest scales. Some of the dependencies on the strength of the
magnetic field, the power law index $n_B$ of the magnetic power
spectrum, the small scale cut-off, and the offset of the observer
position from the middle-plane, are discussed below.

\begin{figure}[ht]
\centering\leavevmode
\epsfxsize=3.5in
\epsfbox{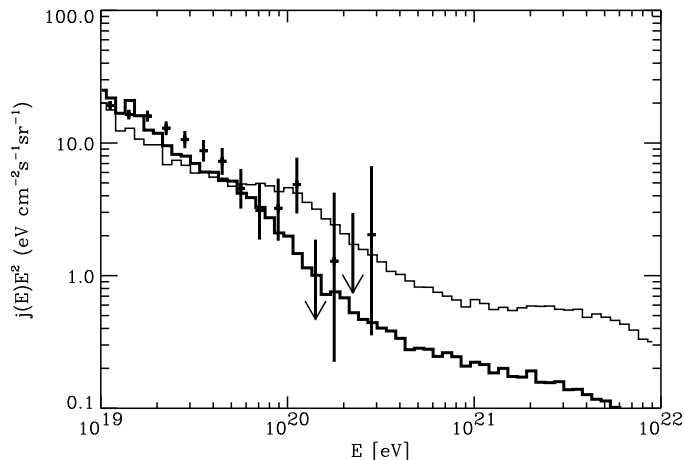}
\caption[...]{Best fit to the combined data above 10 EeV from the
Haverah Park~\cite{Haverah}, Fly's Eye~\cite{FE}, and
AGASA~\cite{AGASA} experiments (error bars) of the
spectra predicted by the diffuse source model explained
in the text. The thick histogram is for $B_{\rm rms}=0.5\,\mu\,$G,
with the observer $2\,$Mpc above the Supergalactic plane,
and the thin histogram is for $B_{\rm rms}=0.05\,\mu\,$G, with
the observer in the plane center.
In both cases the spectra were averaged over 4 magnetic field
realizations with 20000 particles each.}
\label{F1}
\end{figure}

For a diffuse source distribution, any value of the magnetic field
strength, between $\sim0.05\,\mu$G and $\sim0.5\,\mu$G can provide a
reasonable fit to the observed energy spectrum, for a single power-law
injection spectrum $\propto E^{-2.4}$. This is illustrated in
Fig.~\ref{F1}, for the above two extreme values of $B_{\rm rms}$. We
note that the observed energy spectrum is well approximated by a
power-law $\propto E^{-2.7}$ in the range $E\lesssim
100\,$EeV~\cite{AGASA}. The difference in indices between the injected
and the propagated spectrum in Fig.~\ref{F1} results from the diffusion
of UHECRs combined with energy losses~\cite{WW79,SLB99,BO99}; in our
present case, UHECRs with energies greater than $\sim100-1000\,$EeV
propagate in nearly straight lines, in which case the energy spectrum
is unaffected by magnetic deflection, while UHECRs of lower energies
diffuse. The interplay between propagation in the magnetic field and
energy losses leads to a spectrum that is softer than the injection
spectrum in the diffusive regime, but not in the rectilinear regime.
This leads to reasonable fits of the UHECR spectrum above $\simeq10\,$
EeV. We emphasize that limiting the injection to an upper cut-off
$\simeq1000\,$EeV would not change the spectral shape below
$\simeq500\,$EeV.

\begin{figure}[ht] \centering\leavevmode \epsfxsize=3.5in
\epsfbox{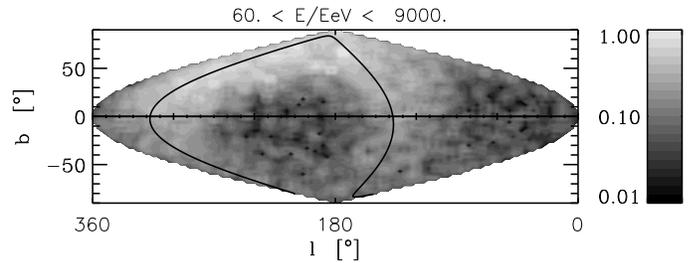} 
 \caption[...]{The angular distribution in Galactic coordinates of
events above 60 EeV, averaged over 4 magnetic field realizations with
20000 particles each for the scenario with $B_{\rm rms}=0.05\,\mu\,$G,
corresponding to the thin histogram in Fig.~\ref{F1}. The grey scale
represents the integral flux per solid angle. The solid line marks the
Supergalactic plane. }
 \label{F2} \end{figure}

\begin{figure}[ht]
\centering\leavevmode
\epsfxsize=3.5in
\epsfbox{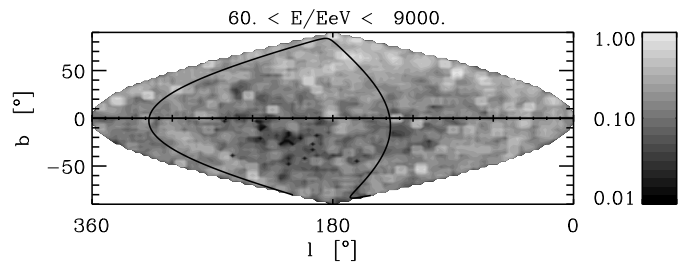}
\caption[...]{Same as Fig.~\ref{F2}, but for the scenario
with $B_{\rm rms}=0.5\,\mu\,$G, corresponding to the thick
histogram in Fig.~\ref{F1}.}
\label{F3}
\end{figure}

The angular distributions associated with these simulations are shown in
Figs.~\ref{F2}, and \ref{F3}, plotted in Galactic coordinates,
for $E>60\,$EeV. These images are averaged over different spatial
realizations of the magnetic field
inside the Local Supercluster. For the case of strong fields
and large coherence lengths, images
corresponding to different realizations are very different from each
other due to cosmic variance, but cover more than half of the
sky, consistent with the isotropy observed by AGASA.

For $B_{\rm rms}\simeq0.05\,\mu$G, shown in Fig.~\ref{F2}, UHECR
arrival directions are strongly clustered around the center of the
Local Supercluster, {\it i.e.} approximately around the Virgo cluster,
at $l\simeq282^\circ$, $b\simeq+75^\circ$. Since AGASA recorded no event
out of 47 for $E\geq40\,$EeV within $\simeq15^\circ$ of this point,
which is furthermore located near the peak of the AGASA exposure curve,
one would exclude this scenario to a high degree of confidence.
However, we note that radio-galaxies in the Supercluster seem to
distribute rather uniformly with radius, {\it i.e.} there is no
preferred center~\cite{SP89}. If the UHECR sources were to follow
radio-galaxies, the sky distribution would be more isotropic, and
further statistics would be needed to discriminate this scenario (see
also below).

For a field strength $B_{\rm rms}\sim0.5 \,\mu$G, the correlation with
the Supergalactic plane disappears, as shown in Fig.~\ref{F3}. This
effect is not trivial, as all UHECR sources in our scenario are located
within the Supergalactic plane.  Note as well that the resulting
angular image in Fig.~\ref{F3} reveals spikes of UHECR events even
though the source is diffuse.  These spikes are produced by magnetic
lensing and give rise to small scale clustering. As we quantify below,
we find that the amplitude of clustering depends on the spectrum of
magnetic inhomogeneities; namely, for a scale-invariant spectrum
$\langle B(k)^2\rangle={\rm constant}$, the spikes are much less
pronounced.  This is expected~\cite{WM96,SLB99}, as the magnetic power
$k^2\langle B(k)^2\rangle$ in that case is concentrated on small
spatial scales, whereas for the Kolmogorov spectrum, the power is
concentrated on large scales.

\begin{figure}[ht] 
\centering\leavevmode 
\epsfxsize=3.5in
\epsfbox{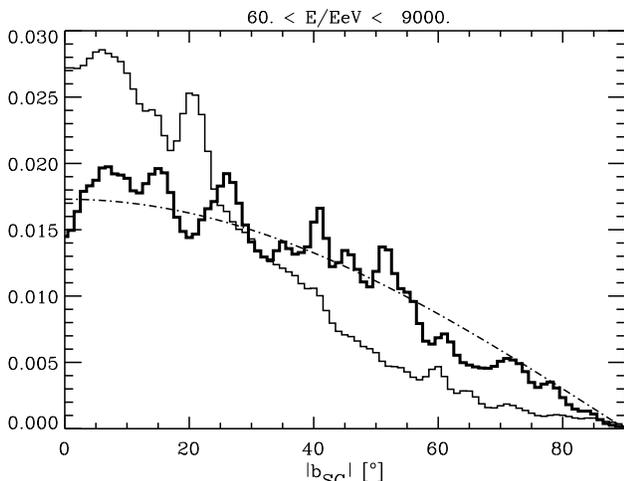} 
\caption[...]{The average over 4 magnetic field realizations of the
distribution of events above 60 EeV in Supergalactic latitude for the
scenario with $B_{\rm rms}=0.5\,\mu\,$G shown in Figs.~\ref{F1}
and~\ref{F3} (thick histogram), and for $B_{\rm rms}=0.05\,\mu\,$G, as
shown in Figs.~\ref{F1} and~\ref{F2} (thin histogram), assuming
$1.6^\circ$ angular resolution. The dash-dotted curve represents a
completely isotropic distribution.} 
\label{F4} 
\end{figure}

In Fig.~\ref{F4}, we show the histogram of events plotted {\it vs.}
Supergalactic latitude; this figure clearly illustrates the relative
anisotropy of a scenario with $B_{\rm rms}=0.05\,\mu$G, with a
distribution of sources strongly centered around Virgo, and the
relative isotropy of a scenario with $B_{\rm rms}=0.5\,\mu$G. A
similar plot for the observed AGASA events is given in
Refs.~\cite{AGASA}.

{\it Observational tests.} A turbulent magnetic field of strength
$B_{\rm rms}\sim0.5\,\mu$G in the Local Supercluster is not
unrealistic~\cite{KCOR97}; it roughly corresponds to the expected
equipartition value $B_{\rm eq}\sim0.2\,\mu{\rm
G}\,\left(T/3\times10^6\,{\rm K}\right)^{1/2}\left(\rho_{\rm
b}/0.3\rho_{\rm c}\right)^{1/2}$, where $T$, $\rho_{\rm b}$, and
$\rho_{\rm c}$ respectively denote the temperature, baryon density and
critical density~\cite{RKB98}. The above scenario thus reopens a window
for conventional astrophysical sources as origin of UHECRs, provided
that such sources can indeed accelerate protons up to
$E\gtrsim1000\,$EeV~\cite{NMA95}. In the following, we give several
predictions and tests of the present scenario.

The most direct test would be to detect and measure a permeating
magnetic field in the Local Supercluster, or at least obtain a trace of
its presence via the observation of synchrotron emission. In this
respect, we note the positive detection of a bridge of synchrotron
emission in the plane of the Coma/Abell~1367 supercluster~\cite{K89},
which lends further support to our scenario. Unfortunately, synchrotron
data does not yield the strength of the magnetic field, and
measurements of Faraday rotation of polarized light from sources
located in the Local Supercluster are needed to this end.

Our model predicts large-scale anisotropy, whose amplitude is expected
to increase with energy above the pion production threshold, because of
decreasing magnetic deflection, and because sources located beyond the
Supergalactic plane, that we neglected here, and that would tend to
make the distribution more isotropic, do not contribute for
$E\gtrsim100\,$EeV~\cite{WFP97}. We quantify this anisotropy in terms
of dipole and quadrupole moments, assuming that there is full sky
coverage, as planned for the Pierre Auger Observatory~\cite{C92}. For
instance, the detection of 64 events above $60\,$EeV, corresponding to
one year of one site of the Pierre Auger Observatory, would allow to
detect anisotropy with a false alarm risk of 5\% (resp. 1\%) in
$77.-99.9$\% (resp.  $63.-99.7$\%) of cases if $B_{\rm
rms}=0.05\,\mu$G, and in $51.-93$\% (resp.  $26.-83.$\%) of cases if
$B_{\rm rms}=0.5\,\mu$G, where the ranges given correspond to variation
of the source distribution, coherence length, $n_B$, and observer
position. These statistics are limited by the finite number or events,
not by cosmic variance. We find that the anisotropy strongly decreases
with increasing radius of the source distribution, and with increasing
magnetic field strength. Finally, we emphasize that a correlation of
UHECR statistics with Supergalactic latitude is not a proper measure of
the predicted anisotropy when $B_{\rm rms}\sim0.5\,\mu$G, because this
correlation starts to differ from the isotropic prediction only for
$E\gtrsim200\,$EeV (see also Fig.~\ref{F4}).

Finally, the observation of a repeated number of clusters of UHECRs
with arrival directions compatible within the angular resolution of
future instruments will also allow to constrain severely the present
model.  A strong magnetic field leads to an almost isotropic arrival
direction distribution on large angular scales. At the same time,
magnetic lensing, illustrated in Ref.~\cite{SLB99} in the case of a
point source, leads to significant small scale clustering. Overall, for
any source distribution the probability for clustering of UHECRs in the
presence of a magnetic field is higher than for an isotropic arrival
direction distribution, and (of course) smaller than in the absence of
magnetic fields with point-like sources.

\begin{table}[ht]
\caption{Probabilities to detect 5 doublets above 40 EeV in
the AGASA data set (47 events), and $\geq1({\rm resp.}\,3)$ multiplets
of $\geq5({\rm resp.}\,3)$ events in a data set of 64 events above 60 EeV,
as expected for 1 year of the Southern Pierre Auger site (PAO),
for $B_{\rm rms}$ as indicated. }
\begin{tabular}{lll}
 $B_{\rm rms}\,(\mu{\rm G})$ & AGASA prob.(\%) & PAO prob.(\%)\\
\tableline
(isotropic) &   0.29    & $\lesssim0.01\,(\lesssim0.01)$ \\
0.05        & $8.-20.$  & $0.3-6.\,(0.5-4.)$ \\
0.5         &  $8.-16.$ & $5.-8.\,(\simeq11.)$ \\
\end{tabular}
\label{T1}
\end{table}

We quantify this in Table~\ref{T1} by comparing clustering
probabilities for our scenario and for a completely isotropic
distribution, in two cases: in the first, we give probabilities
corresponding to the numbers of UHECR events and clusters observed by
the AGASA experiment~\cite{AGASA}. To this end we simulated the finite
angular resolution $\sim1.6^\circ$ and sky coverage of the AGASA
experiment, and calculated the probability of repeated occurences
(multiplets). We call a multiplet a cluster of events such that all
events fall within $2.5^\circ$ of the first event detected. In the
second case, we give an example for numbers of multiplets that would
strongly discriminate our scenario against uniform source distributions
with much weaker magnetic fields, for one year of the Southern Pierre
Auger site, assuming $1^\circ$ angular resolution. The range of values
given corresponds to variation of the source density profile, $n_B$,
coherence length, and position of the observer.

These numbers indicate that the preference for strong fields increases
with exposure. We also found that clustering increases with decreasing
$n_B$, with increasing radius of the source density profile, and with
the coherence length of the field. Within our model, a rather coherent
field, with power on the largest scales, such as a Kolmogorov spectrum
is thus favored. Note that for a less diffuse source distribution with
less abundant but more powerful sources, these probabilities would be
larger.  Our scenario, however, predicts the absence of correlations
between the UHECR clusters and powerful sources associated with the
large scale structure. This can be used to discriminate it also against
models with highly structured source distributions and negligible
magnetic fields~\cite{WFP97}. A somewhat larger number of clusters
and/or higher multiplicity than in Tab.~\ref{T1} would rule out our
scenario, {\it e.g.}, for a cluster of nine events out of 64 showers
above $60\,$EeV, and $B_{\rm rms}=0.5\,\mu$G, the confidence level is
$\lesssim0.1$\%.

{\it Acknowledgments.} We warmly thank the late David Schramm for
constant encouragement and collaboration in earlier work. We
acknowledge P. Blasi and A. Olinto for discussions.  We are grateful to
the Max-Planck Institut f\"ur Physik, M\"unchen (Germany), and the
Institut d'Astrophysique de Paris, Paris (France), for providing CPU
time. This work was supported, in part, by the DoE, NSF, and NASA at
the University of Chicago.


\begin{thebibliography}{9}

\bibitem{RKB98} D.~Ryu and P.~L.~Biermann, Astron.~Astrophys. 335
(1998) 19; see also P.~Blasi, S.~Burles and A.~V.~Olinto,
Astrophys.~J., in press (1999) e-print astro-ph/9812487.

\bibitem{K94} P.~P.~Kronberg, Rep.~Prog.~Phys. 57 (1994) 325.

\bibitem{V97} J.~P.~Vallee, Fund.~Cosm.~Phys. 19 (1997) 1.

\bibitem{WW79} J.~Wdowczyk and A.~W.~Wolfendale, Nature 281
(1979) 356; M.~Giler, J.~Wdowczyk, and A.~W.~Wolfendale, J.~Phys.~G 6
(1980) 1561; V.~S.~Berezinsky, S.~I.~Grigor'eva, and V.~A.~Dogiel,
Zh.~Eksp.~Theor.~Fiz. 96 (1989) 798 [Sov.~Phys.~JETP 69 (1989) 453].

\bibitem{SLB99} G.~Sigl, M.~Lemoine, and P.~L.~Biermann,
Astropart.~Phys. 10 (1999) 141.

\bibitem{BO99} P.~Blasi and A.~V.~Olinto, Phys.~Rev.~D 59 (1999)
023001.

\bibitem{MT98} G.~Medina~Tanco, Astrophys.~J. 505 (1998) L79.

\bibitem{NMA95} J.~P.~Rachen and P.~L.~Biermann, Astron.~Astrophys.
272 (1993) 161; C.~A.~Norman, D.~B.~Melrose, and A.~Achterberg,
Astrophys.~J. 454 (1995) 60; M.~Ostrowski, Astron.~Astrophys. 335
(1998) 1340; E.~Boldt and P.~Ghosh, Mon. Not. Roy. Astron. Soc., in
press (1999) e-print astro-ph/9902342.

\bibitem{AGASA}
M.~Takeda et al., Phys.~Rev.~Lett. 81 (1998) 1163; e-print
astro-ph/9902239, submitted to Astrophys.~J.

\bibitem{C92} J.~W.~Cronin, Nucl. Phys. B (Proc. Suppl.)
28B (1992) 213; The Pierre Auger Observatory Design Report
(2nd ed.) 14 March 1997.

\bibitem{SL98} G.~Sigl and M.~Lemoine, Astropart.~Phys. 9 (1998)
65.

\bibitem{LSO97} M.~Lemoine, G.~Sigl, A.~V.~Olinto, and D.~N.  Schramm,
Astrophys.~J. 486 (1997) L115; G.~Sigl, M.~Lemoine and A.~V.~Olinto,
Phys.~Rev.~D 56 (1997) 4470.

\bibitem{Haverah} M.~A.~Lawrence, R.~J.~O.~Reid, and
A.~A.~Watson, J.~Phys.~G Nucl.~Part.~Phys. 17 (1991) 733.

\bibitem{FE} D.~J.~Bird et al., Phys.~Rev.~Lett. 71 (1993)
3401; Astrophys.~J. 424 (1994) 491; ibid. 441 (1995) 144.

\bibitem{SP89} P.~A.~Shaver and M.~Pierre, Astron.~Astrophys. 220
(1989) 35.

\bibitem{WM96} E.~Waxman and J.~Miralda-Escud\'e, Astrophys.~J. 472
(1996) L89.

\bibitem{KCOR97} R.~M.~Kulsrud, R.~Cen, J.~P.~Ostriker and D.~Ryu,
Astrophys.~J. 480 (1997) 481.

\bibitem{K89} K.~T.~Kim, P.~P.~Kronberg, G.~Giovannini, and T.~Venturi,
Nature 341 (1989) 720; see also T.~A.~Ensslin, P.~L.~Biermann,
U.~Klein and S.~Kohle, Astron.~Astrophys. 332 (1998) 395.

\bibitem{WFP97} E.~Waxman, K.~B.~Fisher, and T.~Piran,
Astrophys.~J. 483 (1997) 1.

\end{thebibliography}
\end{document}